\input harvmac

\def\a{\alpha} \def\b{\beta}  \def\G{\Gamma}
  \def\ee{\varepsilon} 
 \def\th{\theta}  
 \def\l{\lambda}  \def\m{\mu} \def\n{\nu}
\def\cs{\xi}  \def\p{\pi}  \def\r{\rho}
\def\s{\sigma}   
 \def\f{\phi}   
   
\def\pa{\partial}   \def\half{{1\over
2}} 

\Title{}{Duality Transformations Away From Conformal
Points\footnote{$^*$}{This work is supported in part by funds provided
by NSERC of Canada and FCAR of Qu\'ebec. E-mail: {\tt
haagense@cinelli.physics.mcgill.ca}.}}

\centerline{Peter E. Haagensen} \medskip
\centerline{{\it Physics Department, McGill University}}\vskip-.15cm
\centerline{{\it 3600 University St.}}\vskip-.15cm
\centerline{{\it Montr\'eal~~H3A 2T8~~CANADA.}}

\vskip 3cm

\nref\kikkawa{K. Kikkawa and M. Yamasaki, {\it Phys.~Lett.} {\bf 149B}
(1984) 357; N.~Sakai and I.~Senda, {\it Prog.~Theor.~Phys.} {\bf 75}
(1986) 692.}
\nref\alvarez{See for instance, E.~\'Alvarez and M.A.R.~Osorio, {\it
Phys.~Rev.} {\bf D40} (1989) 1150;\phantom{xxxx} E.~Kiritsis, {\it
Nucl.~Phys.}
{\bf B405}
(1993) 109.}
\nref\buscher{T.H.~Buscher, {\it Phys.~Lett.} {\bf 194B}
(1987) 59; {\it Phys.~Lett.} {\bf 201B} (1988) 466.}
\nref\tseytlin{A.A.~Tseytlin, {\it Phys.~Lett.} {\bf 178B}
(1986) 34; {\it Nucl.~Phys.} {\bf B294} (1987) 383.}
\nref\callan{C.G.~Callan, D.~Friedan, E.~Martinec and M.J.~Perry,
{\it Nucl.~Phys.} {\bf B262} (1985) 593.}
\nref\fridling{E.~Braaten, T.L.~Curtright and C.K.~Zachos,
{\it Nucl.~Phys.} {\bf B260} (1985) 630;
B.~Fridling and A.~van de Ven, {\it Nucl.~Phys.} {\bf
B268} (1986) 719.}
\nref\lag{L.~Alvarez-Gaum\'e, D.Z.~Freedman and S.~Mukhi, {\it Ann.~Phys.}
{\bf 134} (1981) 85.}
\nref\hull{C.M.~Hull and P.K.~Townsend, {\it Nucl.~Phys.} {\bf B274}
(1986) 349.}
\nref\lagb{L.~Alvarez-Gaum\'e and D.Z.~Freedman, {\it Phys.~Rev.}
{\bf D22} (1980) 846.}
\nref\tseytlinb{A.A.~Tseytlin, {\it Mod.~Phys.~Lett.} {\bf A6} (1991)
1721.}
\nref\panvel{J.~Panvel, {\it Phys.~Lett.} {\bf 284B} (1992) 50.}
\nref\forgacs{J.~Balog, P.~Forg\'acs, Z.~Horv\'ath and L.~Palla,
{\it ``Perturbative Quantum (In)Equivalence of Dual Sigma
Models in 2 Dimensions"}, {\tt hep-th/9601091}.}

\centerline{{\bf Abstract}}\medskip

Target space duality transformations are considered for bosonic sigma
models and strings away from RG fixed points.  A set of consistency
conditions are derived, and are seen to be nontrivially satisfied at
one-loop order for arbitrary running metric, antisymmetric tensor and
dilaton backgrounds.  Such conditions are sufficiently stringent to
enable an independent determination of the sigma model beta functions at
this order.

\vfill

\bigskip
\noindent McGill-96/14

\noindent {\tt hep-th/9604136}

\Date{04/96}

\newsec{Introduction}\bigskip

Target space duality symmetry (`T-duality', or simply `duality'
henceforth) was first observed for strings compactified on a torus,
where the partition function could be computed and explicitly verified
to be invariant under the transformation $R\rightarrow 1/R$, where $R$
is the target radius \kikkawa.  On these and some other
particular backgrounds, e.g.~WZW
models, duality has in fact been studied in great detail as a symmetry of
the full quantum theory \alvarez. However, even for arbitrary backgrounds
with an abelian isometry, and when the partition function is not known
explicitly, there is a proof of invariance under T-duality
transformations which comes directly from the sigma model path integral
\buscher.  In this proof, all manipulations are done on classical
background fields, and essentially no effect of renormalization is taken
into account.  Duality appears as a classical symmetry, and there is not
really a path integral reasoning connecting it to the quantum properties
of the sigma model.  Naturally, if one starts out with a conformally
invariant background, then it is reasonable to expect that duality will
also be a symmetry at the quantum level, since there will not be any
perturbative quantum corrections to the background.  At one-loop order
this expectation is in fact borne out, and conformally invariant
backgrounds are mapped by classical duality into conformally invariant
backgrounds \buscher.  Nevertheless, conformal backgrounds are a very
small subset of all possible backgrounds, and one might wonder whether
it is possible to extend the action of duality beyond these very special
cases.  While for string theory the main interest may lie in conformal
backgrounds, from a pure 2d field theory point of view, we find it would
be interesting to examine the interplay between duality and RG flows.
This represents the main motivation underlying our investigations here.

Our starting point is a simple and basic observation:  any manipulations
on a path integral can only be meaningful once the path integral itself
is well-defined, and for that one must consider it with proper
regularization and renormalization in place.  The standard manipulations
on classical fields leading to duality transformations, which may be
justifiable for conformal backgrounds, should more appropriately be
considered on the renormalized background fields in general.  Once this
is done, duality transformations will still have the same form, but now
they act on renormalized fields which, in particular, flow with a
renormalization group parameter $\mu$.  As such, duality then maps
entire renormalization flows into other renormalization flows and thus,
like other symmetries in field theory, it also points to the fact that
our parametrization of a certain field space is redundant.  More
importantly, once one considers this flow in the duality
transformations, we will show that a stringent set of consistency
conditions follows on the possible quantum corrections in order that
they respect duality symmetry.  These conditions are expressed as linear
homogeneous relations among the beta functions of the original and dual
theories.

We will investigate the validity of these consistency relations in all
generality at one-loop order.  The result we find is that they are
satisfied for entirely generic backgrounds of metric, antisymmetric
tensor and dilaton, due to the specific form the beta functions take.
In fact, as we shall see explicitly, one can turn the argument around
and actually {\it derive} the one-loop beta functions simply by
requiring the consistency relations.

An important distinction should be made between the sigma model on a
flat worldsheet (or simply `sigma model' in what follows) and the
string,
in order to make more precise what we have been loosely referring to as
`beta functions' in the above.  While for the sigma model the quantities
appearing in the consistency relations are the beta functions, in the
string, there is an extra background field, the dilaton, and the
relevant quantities are the Weyl anomaly coefficients \tseytlin.  In
this latter case, the consistency relations will only hold provided the
well-known dilaton shift \buscher\ is also implemented.  Apart from
this,
the difference that occurs between the two cases is that the relations
will be exactly satisfied for the Weyl anomaly coefficients, while the
sigma model beta functions will only satisfy them after a particular
field redefinition.  At conformal points of the string, in particular,
this will imply the well-known statement that conformal backgrounds are
mapped to conformal backgrounds under duality.  The analogous statement
for the sigma model is that scale invariant (or on-shell finite)
backgrounds will be mapped to scale invariant backgrounds.

In what follows, we will initially write down the sigma model under
consideration, namely, on a generic metric and antisymmetric tensor
background, and the duality transformations that ensue from the abelian
isometry being assumed for the model.  We will then derive the
consistency relations that follow for the quantum corrections to the
background.

Both the duality transformations and the consistency relations treat the
background in a way which breaks manifest target space covariance.  In
order to deal with this, we will verify the consistency relations
through a Kaluza-Klein decomposition of the background tensors.  We will
present the well-known results for the one-loop beta functions and Weyl
anomaly coefficients \tseytlin, and then proceed to perform the
decomposition on these quantities.  When this is done, it is then a
lengthy but straightforward exercise to show that the relations are
indeed satisfied without restrictions on the background.  We will
explicitly see how the dilaton shift re-emerges in this context, as well
as the specific $O(\a ')$ field redefinition necessary for the sigma
model beta functions.  It will also become clear that the mixing of
torsion and geometry entailed by duality is exactly ``matched'' by the
precise admixture of torsion and geometry in the one-loop beta
functions.  This is what allows for an independent calculation of the
beta functions at this order starting from our consistency requirements.

\newsec{Duality Transformations and Consistency Relations}\medskip

Our starting point is the $d\!=\!2$ bosonic sigma model in a generic
$D\!+\! 1$-dimensional background $\{ g_{\m\n}(X),b_{\m\n}(X)\}$ of
metric and antisymmetric tensor, respectively, where $\m,\n =0,1,\ldots
,D=0,i$, so that the $\m\!  =\! 0$ component is singled out.  We shall
assume this sigma model has an abelian isometry, which will enable
duality transformations, and we shall consider the background above in
the adapted coordinates, in which the abelian isometry is made manifest
through independence of the background on the coordinate $\theta\equiv
X^0$ \buscher.  The original sigma model action reads:
\eqn\original{\eqalign{S={1\over 4\p\a '}\int d^2\!\s\, &\left[
g_{00}(X) \pa_\a \th\pa^\a\th +2g_{0i}(X)\pa_\a\th\pa^\a
X^i+g_{ij}(X)\pa_\a X^i\pa^\a X^j+ \right.\cr &\left. i\ee^{\a\b}\left(
2b_{0i}(X)\pa_\a\th\pa_\b X^i+b_{ij}(X)\pa_\a X^i \pa_\b
X^j\right)\right]\, .}}
Not only in the above, but throughout, all background tensors can depend
only on target coordinates $X^i$, $i=1,\ldots ,D$, and not on $\th$.

Regularization and renormalization of the sigma model is
achieved typically through dimensional regularization plus background
field method plus normal coordinate expansion \lag. Upon such regularization
a quantum effective action is obtained, in which the background metric
and antisymmetric tensor become functions of a renormalized scale $\m$ through
quantum corrections in the form of curvature and torsion terms. Such a
procedure is standard and well-known, and we will assume it here {\it ab
initio}.

The duality transformations in this model are also well-known \buscher:
\eqn\duality{\eqalign{\tilde{g}_{00}&={1\over g_{00}}\cr
\tilde{g}_{0i}&={b_{0i}\over g_{00}}\ ,\ \ \ \tilde{b}_{0i}={g_{0i}\over
g_{00}}\cr \tilde{g}_{ij}&=g_{ij}-{g_{0i}g_{0j}-b_{0i}b_{0j}\over g_{00}}\cr
\tilde{b}_{ij}&=b_{ij}-{g_{0i}b_{0j}-b_{0i}g_{0j}\over g_{00}}\, .}}

The statement of classical duality is that the model defined on the
dual background $\{\tilde{g}_{\m\n},\tilde{b}_{\m\n}\}$ is simply a
different parametrization of the same model, given that the
manipulations used to derive the transformations essentially only
involve performing trivial integrations in a different order starting
from the path-integral in which the abelian isometry is gauged.

Our further observation here, which represents the starting point of
this investigation, is that such manipulations must be considered on a
properly regularized path integral, with bare background fields
containing all the necessary counterterms {\it in lieu} of the classical
fields (one should also assure oneself that once the classical fields
have a certain isometry manifest in a certain coordinate system, so will
the bare fields including all perturbative corrections.  A moment's
thought shows this is generically true).  Once this is done, the usual
procedure of interchanging the order of integrations in the gauged model
\buscher\ will lead to two
quantum effective actions, one based on the original background, and one
based on the dual one.  The duality transformations remain the same as
the ones above, but all quantities should be taken to be a function of a
renormalization scale $\m$, so that their flow under changes in $\m$
succintly take into account the quantum corrections they contain.
Denoting by
\eqn\betas{\b_{\m\n}^g\equiv \m {d\over d\m}g_{\m\n}\,
,\,\,\, \b_{\m\n}^b\equiv \m {d\over d\m}b_{\m\n}}
the metric and
antisymmetric tensor beta functions, respectively, and applying $\m
d/d\m$ to the duality transformations, one obtains:
\eqn\consistency{\eqalign{ \b^{\tilde{g}}_{00}&=-{1\over g_{00}^2}
\b^g_{00}\cr \b^{\tilde{g}}_{0i}&=-{1\over g_{00}^2}\left(
b_{0i}\b^g_{00}-\b^b_{0i}g_{00} \right) \cr
\b^{\tilde{b}}_{0i}&=-{1\over g_{00}^2}\left(
g_{0i}\b^g_{00}-\b^g_{0i}g_{00} \right)\cr
\b^{\tilde{g}}_{ij}&=\b^g_{ij}-{1\over g_{00}}\left( \b^g_{0i}g_{0j}+
\b^g_{0j}g_{0i}-\b^b_{0i}b_{0j}-\b^b_{0j}b_{0i}\right) + {1\over
g_{00}^2}\left( g_{0i}g_{0j}-b_{0i}b_{0j}\right) \b^g_{00}\cr
\b^{\tilde{b}}_{ij}&=\b^b_{ij}-{1\over g_{00}}\left( \b^g_{0i}b_{0j}+
\b^b_{0j}g_{0i}-\b^g_{0j}b_{0i}-\b^b_{0i}g_{0j}\right) + {1\over
g_{00}^2}\left( g_{0i}b_{0j}-b_{0i}g_{0j}\right) \b^g_{00}\, ,}} where
the quantities on the l.h.s. are the beta functions of the dual
background. These are then the consistency relations that the beta
functions must satisfy in order that quantum corrections in the sigma
model satisfy classical duality symmetry.  The
above conditions may also be seen as a statement of ``covariance" of the
renormalization group flow under duality transformations.  It should be
noted that the above relations imply highly nontrivial and restrictive
constraints, since classical duality transformations have in principle
no information on what the actual renormalization of the theory is.

On a generic background, the one-loop beta functions are
\callan,\fridling,\tseytlin :
\eqn\betaone{\eqalign{\b_{\m\n}^g=&\a '\left( R_{\m\n} -{1\over4}
H_{\m\l\r} H_{\n}^{\ \l\r}\right)\cr \b_{\m\n}^b=&-{\a '\over2}\nabla_\l
H^\l_{\ \m\n}\ ,}} where $H_{\m\n\l}=\pa_\m b_{\n\l}+{\rm cyclic\
permutations}$, and $\nabla_\m$ denotes the torsionless covariant
derivative.  We must verify that these beta functions satisfy the
conditions above if the original and dual backgrounds are related as in
\duality . We will also consider the analogous relations for the Weyl
anomaly coefficients, which are given at one loop by \tseytlin:
\eqn\weyl{\eqalign{\bar{\b}_{\m\n}^g=&\b_{\m\n}^g+2\a '\nabla_\m
\pa_\n\f\cr \bar{\b}_{\m\n}^b=&\b_{\m\n}^b+\a 'H_{\m\n}^{\ \
\l}\pa_\l\f\ .}} \medskip

\newsec{Kaluza-Klein Decomposition}\medskip

In order to verify identities which break manifest target space
covariance in one direction, we find that the most direct and economical
way to proceed is to consider the decomposition of tensors which is
typical of Kaluza-Klein reductions.  We write the arbitrary metric
$g_{\m\n}$ as:
\eqn\metric{g_{\m\n}=\pmatrix{a & av_i\cr av_i &
\bar{g}_{ij} +av_iv_j}\ ,} so that $g_{00}\!=\!a, g_{0i}\!=\!av_i,
g_{ij}\!=\!\bar{g}_{ij}+av_iv_j$, and all quantities do not depend on
$X^0\!=\!\th$.  The components of the antisymmetric tensor are also
decomposed as $b_{0i}\equiv w_i$ and $b_{ij}$.  Under \duality, the dual
background is easily found to be:
\eqn\dualmetric{\tilde{g}_{\m\n}=\pmatrix{1/a & w_i/a\cr w_i/a &
\bar{g}_{ij}+w_iw_j/a}\ ,} and $\tilde{b}_{0i}\!=\!
v_i,\tilde{b}_{ij}\!=\!b_{ij}+w_iv_j-w_jv_i$.

We now need to work out the expression for the connection coefficients
and Ricci tensor for both original and dual geometries, but of course we
only need to do it once, since the dual geometry is obtained from the
original one by the substitution $a\rightarrow 1/a, v_i\rightarrow w_i$.
Likewise, dual torsion is obtained from the original one by
$w_i\rightarrow v_i$ and $b_{ij} \rightarrow b_{ij}+w_iv_j-w_jv_i$.  We
list below all geometric quantities relevant for our computation:
\item{1)} {\it inverse metric}:  $g^{00}\!=\!1/a+v_iv^i,\
g^{0i}\!=\!-v^i,\ g^{ij}\!=\!\bar{g}^{ij}$.  On decomposed tensors,
indices $i,j,\ldots$ are raised and lowered with the metric
$\bar{g}_{ij}$ and its inverse.  We note also that $\det g= a
\det\bar{g}$.

\item{2)} {\it connection coefficients}:
\eqn\connection{\eqalign{
\G^0_{00}&={a\over2}v^ia_i\ ,\ \G^0_{i0}={a\over2}\left[ {a_i\over a}+
v^ja_jv_i+v^jF_{ji}\right]\cr
\G^i_{00}&=-{a\over2}a^i\ ,\ \G^i_{0j}=-{a\over2}\left[ F^i_{\
j}+a^iv_j\right]\cr
\G^0_{ij}&=-\bar{\G}^k_{ij}v_k+\half (\pa_iv_j+\pa_jv_i+a_iv_j+a_jv_i)-
{a\over2}v^k\left[ v_jF_{ik}+v_iF_{jk}-a_kv_iv_j\right]\cr
\G^i_{jk}&=\bar{\G}^i_{jk}+{a\over2}\left[ v_jF_k^{\ i}+v_kF_j^{\ i}
-a^iv_jv_k\right]\ ,}}
where $a_i\!=\!\pa_i\ln a\ ,\ F_{ij}\!=\!\pa_iv_j-\pa_jv_i$, and
$\bar{\G}^i_{jk}$ are the connection coefficients for the metric
$\bar{g}_{ij}$.

\item{3)} {\it Ricci tensor}:
\eqn\ricci{\eqalign{
R_{00}&=-{a\over2}\left[ \bar{\nabla}_ia^i+\half a_ia^i-{a\over2}F_{ij}
F^{ij}\right]\cr
R_{0i}&=v_iR_{00}+{3a\over4}a^jF_{ij}+{a\over2}\bar{\nabla}^jF_{ij}\cr
R_{ij}&=\bar{R}_{ij}+v_iR_{0j}+v_jR_{0i}-v_iv_jR_{00}-\half \bar{\nabla}_ia_j
-{1\over4}a_ia_j-{a\over2}F_{ik}F_j^{\ k}\ ,}}
where, again, barred quantities refer to the metric $\bar{g}_{ij}$.

\item{4)} {\it torsion}:
\eqn\torsion{\eqalign{ H_{0ij}&=-\pa_iw_j+\pa_jw_i\equiv -G_{ij}\cr
H_{ijk}&=\pa_ib_{jk}+\pa_jb_{ki}+\pa_kb_{ij}\ ,}} and all other
components vanish.  For the metric beta function we need
\eqn\torsionsq{\eqalign{ H_{0\m\n}H_0^{\ \m\n}&=G_{ij}G^{ij}\cr
H_{0\m\n}H_i^{\ \m\n}&=-2G_{ij}G^{jk}v_k-H_{ijk}G^{jk}\cr
H_{i\m\n}H_j^{\ \m\n}&=2\left({1\over a}+v_mv^m\right) G_i^{\ k}G_{jk}
-2v^kv^mG_{ik}G_{jm}+2\left( H_{ikm}G_j^{\ k}v^m+i\leftrightarrow
j\right)\cr &\ +H_{ikm}H_j^{\ km}\ ,}} and for the antisymmetric tensor
beta function we need
\eqn\divtorsion{\eqalign{ \nabla_\m H^\m_{\
0i}&=\bar{\nabla}^jG_{ji}-aG_{ij}F^{jk}v_k+\half G_{ij}a^j
-{a\over2}F^{jk}\left( H_{ijk}+v_iG_{jk}\right)\cr \nabla_\m H^\m_{\
ij}&=\bar{\nabla}^k\left( H_{kij}+v_kG_{ij}\right) -\half \left[ G_i^{\
k}\bar{\nabla}_{(k}\ v_{j)} -G_j^{\ k}\bar{\nabla}_{(k}\ v_{i)}
\right]\cr &-{a\over2}v_{[i}H_{j]km}F^{km}+v_{[i}G_{j]k}\left(
a^k-aF^{km}v_m\right)\cr &+\half a^kH_{kij}+\half v_ma^mG_{ij}-\half
F_{[i}^{\ k}G_{j]k}\ ,}} where $[ij]=ij-ji$ and $(ij)=ij+ji$.

\item{5)} {\it dilaton terms}:
\eqn\dilaton{\eqalign{
\nabla_0\pa_0\f&={a\over2}a^i\pa_i\f\cr
\nabla_0\pa_i\f&={a\over2}\left(F^j_{\ i}+a^jv_i\right) \pa_j\f\cr
\nabla_i\pa_j\f&=\bar{\nabla}_i\pa_j\f -{a\over2}\left( v_iF_j^{\ k}+
v_jF_i^{\ k}-a^kv_iv_j\right)\pa_k\f\ .}}
For the dual background, we take $\f\rightarrow\tilde{\f}$, with $\tilde{\f}$
as yet undetermined. Consistency conditions will tell us what it is.

One must also compute the analogous quantities in the dual background,
through the substitutions $a\rightarrow 1/a\ ,\ v_i \leftrightarrow w_i\
,\ b_{ij}\rightarrow b_{ij}+w_iv_j-w_jv_i$.  This is straightforward,
and we will not do it here. These results (and quite some patience!) are
essentially all one needs to verify consistency relations
\consistency\ at one-loop order both for Weyl anomaly coefficients and
beta functions. We show this explicitly for the first and simplest
one (the $00$ component), since it already presents all the essential
features we have alluded to above. Starting with the Weyl anomaly
coefficient:
\eqn\weylzero{\eqalign{
\bar{\b}^{\tilde{g}}_{00}=&\tilde{R}_{00}-{1\over4}\tilde{H}_{0\l\r}
\tilde{H}_0^{\ \l\r}+2\tilde{\nabla}_0\pa_0\tilde{\f}\cr
=&-{1\over2a}\left( -\bar{\nabla}_ia^i+\half a_ia^i-{1\over2a}G^2
\right) -{1\over4}F^2-{1\over a}a^i\pa_i\tilde{\f}\cr
-{1\over g_{00}^2}
\bar{\b}^{g}_{00}=&-{1\over g_{00}^2}\left(
{R}_{00}-{1\over4}{H}_{0\l\r}
{H}_0^{\ \l\r}+2{\nabla}_0\pa_0{\f}\right)\cr
=&-{1\over a^2}\left[ -{a\over2}\left(\bar{\nabla}_ia^i+\half
a_ia^i-{a\over2}F^2
\right) -{1\over4}G^2+ aa^i\pa_i{\f}\right]\ ,}}
with $G^2\equiv G_{ij}G^{ij}$ and $F^2\equiv F_{ij}F^{ij}$ (we have set
$\a '\!=\!1$ since it is unimportant here). These two
expressions will only match once we make the identification $\tilde{\f}
=\f-\half\ln a$. This reproduces the well-known dilaton shift present in
duality transformations.

The analogous relation for the beta function is:
\eqn\betazero{\eqalign{
{\b}^{\tilde{g}}_{00}=&\tilde{R}_{00}-{1\over4}\tilde{H}_{0\l\r}
\tilde{H}_0^{\ \l\r}\cr
=&-{1\over2a}\left( -\bar{\nabla}_ia^i+\half a_ia^i-{1\over2a}G^2
\right) -{1\over4}F^2\cr
-{1\over g_{00}^2}
{\b}^{g}_{00}=&-{1\over g_{00}^2}\left( {R}_{00}-{1\over4}{H}_{0\l\r}
{H}_0^{\ \l\r}\right)\cr
=&-{1\over a^2}\left[ -{a\over2}\left(\bar{\nabla}_ia^i+\half
a_ia^i-{a\over2}F^2
\right) -{1\over4}G^2\right]\ ,}}
(with $\a '\!=\!1$ here again). We immediately realize that since the
dilaton is not present, neither is the dilaton shift which was necessary
previously for consistency, and the two
expressions above do not match. However, an $O(\a ')$ field redefinition
coming from a target reparametrization \hull\ in the original model
cures this mismatch:
\eqn\redefg{{\b}^{g}_{\m\n}\rightarrow{\b}^{'g}_{\m\n}=
{\b}^{g}_{\m\n}-\a '\nabla_{(\m}\ \cs_{\n )}\ ,}
with $\cs_\m=-1/2\ \pa_\m\ln a$, so that
\eqn\betaprimeg{{\b}^{'g}_{00}={\b}^{g}_{00}+{\a '\over2}a a_ia^i
\Rightarrow
{\b}^{\tilde{g}}_{00}=-{1\over g_{00}^2}{\b}^{'g}_{00}\ ,}
so that consistency is again verified. It is worthwhile noting that this
fits nicely with the fact that while the string has a dilaton and the
sigma model does not, the Weyl anomaly coefficients do not transform
under field redefinitions, but the beta functions do \tseytlin. An analogous
mismatch will occur in the relations in which the antisymmetric tensor
beta function is involved. It will again be removed with the same field
redefinition as above, where now the antisymmetric beta function also
changes to
\eqn\redefb{{\b}^{b}_{\m\n}\rightarrow{\b}^{'b}_{\m\n}=
{\b}^{g}_{\m\n}-\a 'H_{\m\n}^{\ \ \l}\cs_{\l}}
up to a gauge transformation.  Recalling that scale invariance of the
sigma model (which is equivalent to on-shell finiteness) requires that
the beta functions be
\eqn\finite{\eqalign{{\b}^{g}_{\m\n}&=\nabla_{(\m}\ V_{\n )}\cr
{\b}^{b}_{\m\n}&=H_{\m\n}^{\ \ \l} V_{\l}\ ,}}
for $V_\m$ some target vector, the above then shows that under classical
duality a one-loop finite sigma model will be mapped to a one-loop
finite sigma model.

We finally note that while the $G^2$ term in \betazero\ comes from
geometry in the dual model, it comes from torsion in the original one,
and vice-versa for the $F^2$ term.  On the other hand, we know from scaling
arguments \lagb\lag\ that a one-loop metric beta function must be a linear
combination of the geometric quantities $R_{\m\n}$, ${H}_{\m\l\r}
{H}_\n^{\ \l\r}$, $g_{\m\n}R$ and $g_{\m\n}H_{\l\r\s}H^{\l\r\s}$.
Because of this ``interchange" of torsion
and geometry, if we take an arbitrary linear combination of these four
terms for the metric beta function and require the above consistency
conditions, we find a relative factor $-1/4$ between the first two terms,
while the last two terms, $g_{\m\n}R$ and
$g_{\m\n}H_{\l\r\s}H^{\l\r\s}$, have vanishing coefficients
(this
is found already for the $00$ component).\footnote{$^\dagger$}{We thank
D.Z.~Freedman and R.C.~Myers for reminding us that scaling arguments
do not rule out the terms $g_{\m\n}R$ and
$g_{\m\n}H_{\l\r\s}H^{\l\r\s}$ from the one-loop metric beta
function.}   This same interchange of
torsion and geometry between the original and dual backgrounds will take
place in all of equations \consistency\ (but in a considerably more
involved way than simply $G^2 \leftrightarrow F^2$ for the other
relations).  The other relations contain furthermore terms coming from
$R_{\m\n}$ matching terms coming from $\nabla_\l H^\l_{\ \m\n}$.  This
matching fixes their relative factor of $-1/2$.  Antisymmetry in $\m\n$
and the same scaling arguments lead to this latter quantity
as the only possible one-loop antisymmetric beta function, and
altogether then, the one-loop metric and antisymmetric tensor beta
functions are completely determined up to a global constant (which of
course cannot be determined from these relations since they are linear
and homogeneous).

Components $0i$ of \consistency\ require the matching of roughly 20
different terms, while components $ij$ involve approximately 80
different terms. We will not present any details of these calculations
here, but rather just mention that they exactly corroborate the claims
we have made through examination of the $00$ component alone.\medskip

\newsec{Conclusions}\medskip

In this Letter we have considered the action of T-duality
transformations on backgrounds away from conformal points in order to
study the interplay between classical duality symmetry and RG flows.  A
set of consistency relations follow for the quantum corrections to the
sigma model which are satisfied in all generality at one-loop order,
showing that RG flows are entirely compatible with duality at this
order.  At limiting points of the flows, the consistency relations state
that for the sigma model, scale invariant backgrounds are mapped to
scale invariant backgrounds, while for the string, the well-known
statement that conformally invariant backgrounds are mapped to
conformally invariant backgrounds is recovered.  These consistency
relations are stringent enough to allow for an independent determination
of the one-loop beta functions up to one global factor.

There have been to date a few investigations on the issue of
preservation of classical duality symmetry on conformal backgrounds at
two-loop order, and possible corrections to the classical
transformations \tseytlinb\panvel\forgacs.  Ref.  \tseytlinb, in
particular, dealing with the more restricted case of torsionless
original and dual backgrounds, proposes a correction to the
transformations in order to preserve duality symmetry of the two-loop
string low energy effective action.  Such corrections probably indicate
subtleties regarding duality transformations which are not entirely
accounted for in the standard path integral derivation in \buscher.  Any
such perturbative modifications of the transformations at the conformal
point are likely to engender modifications in the transformations on the
running couplings as well, and thus on the consistency relations which
follow from these.  It would be interesting to examine consistency
relations in the presence of such modifications.  Such investigations,
based mainly on cases treated in \tseytlinb\ and \forgacs, are currently
in progress.\bigskip

{\bf Acknowledgments}

It is a pleasure to thank Peter Forg\'acs for the discussions which got
me initiated in this project.  I am also very grateful to Nemanja
Kaloper and Rob Myers for innumerable discussions on T-duality.

\listrefs

\bye